\documentstyle[12pt]{article}

\begin{document}
\title{\vspace*{-1cm}Nuclear Matter with Quark-Meson Coupling I:\\
Comparison of Nontopological Soliton Models}
\author{
Nir Barnea\footnote{e-mail: barnea@ect.it} and 
Timothy S. Walhout\footnote{e-mail: walhout@ect.it} \\
{\small \it ECT*, European Centre for Theoretical Studies in Nuclear Physics 
and Related Areas,}\\
{\small \it Strada delle Tabarelle 286, I-38050 Villazzano (Trento), Italy}\\
{\small and}\\
{\small \it Istituto Nazionale di Fisica Nucleare, Gruppo
collegato di Trento}\\
\vspace {-0.6mm}
        }
\maketitle

\setcounter{page}{1}

\maketitle
\begin{abstract}

A system of nontopological solitons interacting through scalar and vector
meson exchange is used to model dense nuclear matter. The models
studied are of the
Friedberg-Lee type, which exhibit dynamical bag formation due
to the coupling of quarks to a scalar composite gluon field $\sigma$.
It is shown 
that the high density behavior of such models in the Wigner-Seitz
approximation depends essentially
on the leading power of the quark-$\sigma$ coupling vertex. 
By insisting that the parameters of any soliton model be chosen
to reproduce single nucleon properties, 
this high-density behavior then selects a promising class
of models that better fit the empirical results --- the 
chiral chromodielectric models. 
It is shown that one can adjust
the model to obtain saturation as well as an increase 
of the proton charge radius with nuclear density. 
These two phenomena depend on
the presence of a scalar meson and were not found in other nuclear
matter calculations based on Friedberg-Lee type models.
\end{abstract}

\setlength{\parskip}{0.0in}



\hspace*{2.03cm}{\small Model, Nuclear Matter}

\setlength{\textheight}{9.0in}
\setlength{\parskip}{0.1in}
\newpage


\section{Introduction}

\hspace* {6mm}

Ever since the advent of quantum chromodynamics (QCD) it has been
popular to describe the nucleon in terms of bag or soliton models.
There are many versions of such models, characterized by two extremes:
the MIT bag model \cite{MIT}, where the nucleon consists of just constituent
quarks arbitrarily restricted in a given volume, and the
Skyrme model \cite{Skyrme}, where there are no quarks and the nucleon is
instead a topological soliton of the pion field. Other
models interpolate between these two extremes, seeking to combine
the obvious emphases on the structure of nucleons 
within the MIT bag model and on nuclear interactions
via meson exchange within the Skyrme model.
For example, chiral bag models \cite{CBM} surround an MIT-like
bag of quarks with a Skyrme-like cloud of pions (and perhaps
other mesons as well). Clearly, each such
model attempts to balance between low-energy degrees of freedom ---
the mesons --- and high-energy degrees of freedom --- the quarks (and possibly
gluons as well). Our aim here is to develop a model that as simply
as possible, but without losing essential physics, 
combines quark and meson degrees of freedom. This task is made more
difficult by the ``Cheshire cat principle'' \cite{Cheshire}, 
an extrapolation of results found from chiral bag models, 
which find that low-energy
nucleon properties are largely insensitive to the size of the
quark bag. As the bag shrinks, the meson cloud forms more and
more of the nuclear structure.  To select a model, we must study
nuclear properties that distinguish between a bag of quarks and
a cloud of mesons.

An obvious testing ground for any bag or soliton model is dense
baryon matter, where the structure of individual nucleons will
become as important as the interactions between neighboring nucleons.
Preferably, the model will have a dynamical formation
of the bag or soliton, for then one can treat the transition to
a quark-gluon plasma consistently. Surprisingly enough, the
Skyrme model, which may be thought of as a chiral bag model with
the quark bag shrunk to zero size, does predict a dynamical transition
to a phase of solitons of fractional baryon number at high
densities \cite{Wal88}; however, identifying this phase with 
a quark-gluon plasma is certainly problematic.
Thus in this paper we study dense
nuclear matter in a set of models known 
as non-topological soliton (NTS) models.
In particular, we study the Friedberg-Lee (FL) model 
\cite{Fri77} and a class of
related models, the chiral chromodielectric ($\chi$CD) models
\cite{Fai88,Wil89}, 
as well as
extensions of these models (or, one might argue, approximations
to these models) that explicitly include couplings to mesons.
These models, based upon general arguments from QCD, are
characterized by the coupling of quarks to a scalar field $\sigma$ that
has a nonzero vacuum expectation field. This field is understood
to be a composite gluon field. The interaction between the quarks
and scalar field leads to a dynamical confinement mechanism, with
the quarks carving a hole in the background scalar field. The
structure of this bag depends precisely on how the quarks couple
to the scalar field, and it is this coupling that distinguishes
the various models we study here. All these models reproduce
single nucleon properties reasonably well --- indeed, chiral
bag models show that single nucleon properties are relatively
insensitive to the structure of the quark bag. We must look
at high densities, where the bags begin to overlap,
to see the differences between models with different quark-$\sigma$
couplings.

In the last few years there also has been considerable interest in 
the application of the quark-meson coupling (QMC) model to 
study the nuclear matter equation of state, as well as medium effects on the
nucleon structure and nucleon-meson coupling,
motivated by the  
apparent success of the Walecka QHD models \cite{Wal74,Ser86} 
in describing the properties of nuclear matter.
The QMC model, initially suggested by Guichon
\cite{Gui88}, consists of non-overlapping nucleon (MIT) bags
that interact through the exchange of scalar 
and vector mesons in the mean-field approximation (MFA).
This model has been generalized to include Fermi motion and center of mass
corrections \cite{Fle90} and was applied to nuclear matter (see \cite{Mul97}
and references therein)
and also to finite nuclei \cite{Gui96,Blu96}.  
Of course, the assumption that the nucleons can 
be regarded as non-overlapping bags is only valid at low density, where
these models seem to capture the essential physics. 
Already at nuclear saturation density, however,
the internucleon separation is
comparable to the nucleon radius, and at higher densities the assumption
that the bags do not overlap clearly breaks down.
This assumption becomes even more questionable when a modification
of the bag constant in nuclear matter is taken into account.
Jin and Jennings \cite{Jin97} and M\"uller and Jennings \cite{Mul97}
 have recently shown that the introduction of a density-dependent
bag constant can
reproduce the EMC effect and reconcile the QMC results with those of
the Walecka QHD-I model \cite{Wal74}. As a result \cite{Mul97}
the bag radius grows with increasing density and the
overlapping of the bags starts just above nuclear saturation density. 

To take into account the effects from overlapping nucleon bags --- 
that is, to study a nuclear liquid rather a nuclear gas --- it
is clearly of interest to
introduce some dynamics in the confining mechanism. Thus we
are led to replace the MIT bag model by a
NTS model. Although the study of nuclear matter
properties was begun long ago (see \cite{Wil89} and references therein)
and was carried out with different soliton models and with different
levels of sophistication (see \cite{Bir88,Web97} and 
references therein), none of these
calculations included the effect of background meson fields on
nuclear matter. Now, in principle, the non-topological soliton models 
--- in particular, the $\chi$CD model, which in its full form includes
perturbative gluon exchange --- contain sea quarks and meson
exchange explicitly. However, in practice it is very difficult to
deal with anything besides the nuclear constituent quarks. In the
interest of simplicity (and at the sacrifice of consistency), we
extend our NTS models to include explicitly meson degrees of
freedom. (Such an extension has been referred to as the
local uniform approximation to $\chi$CD in \cite{Kre91}.)

In the following, then, we shall study a system of
non-topological solitons interacting via
the exchange of scalar and vector mesons within the MFA.
In modeling nuclear matter each soliton is centered in a 
spherical Wigner-Seitz (WS) cell. It has been argued
that the choice of a spherical WS cell is more appropriate
for a fluid phase
than a crystal as it represents an angular average over neighbouring
sites. We consider two further approximations: in one, the WS 
calculation is simply used to provide an effective nucleon
mass and the kinetic energy is then
taken to be that of a Fermi gas, thus
providing the correct low-density limit. 
A second approximation uses a Bloch-like boundary condition
to calculate the band structure of the quark states.
Neither of these approximations is quite satisfactory, and
the second paper of this series is devoted to improving the
modeling of a liquid of solitons. Nevertheless,
we can still hope that the major qualitive features
of dense nuclear matter will be reproduced even given the
two approximations used here. 
Indeed, we find some
encouraging results such as nuclear saturation and an increase
of the proton rms with nuclear density. These effects are shown to derive
from the background meson fields, which have a considerable
effect on the formation of energy bands in nuclear matter.

In this paper we limit our attention to Friedberg-Lee soliton type models
extended to include the meson fields. The non-topological soliton models
are presented in Sec. \ref{sec:models}.
The Wigner-Seitz approximation is then discussed in Sec. \ref{sec:WS}.
In Sec. \ref{sec:trivial} we study in some detail the trivial solution
of the model. Using these solutions we show
that the high density limit of the Friedberg-Lee type models
depends on the leading power of the quark-$\sigma$ coupling vertex. 
The numerical results are presented in Sec. \ref{sec:results}

\section{Soliton models with quark-meson coupling} 
\label{sec:models}

Our starting point for studying the high density behavior of soliton
matter is the Friedberg-Lee \cite{Fri77} non-topological 
soliton model, and we also consider
related models like the chiral chromo\-dielectric model
of Fai, Perry and Wilets \cite{Fai88}. In their simplest versions,
these models include only constituent quarks
and a single scalar field $\sigma$ that couples to the quarks.
For the light quarks we shall assume $m_u=m_d=0$.
Extending these models in the spirit of the quark-meson coupling 
model, we introduce in addition two meson fields: namely, a scalar
meson $\phi$ and a vector meson $V_{\mu}$, which play important roles
in quantum hadrodynamics. We assume these mesons couple linearly
to the quarks. 
There is some freedom in the structure of the quark-meson vertex, in regard 
to its dependence on the soliton field $\sigma$. Using the Nielsen-Patkos
Lagrangian \cite{Nie82}, Banerjee and Tjon \cite{Ban97} have recently
argued that in NTS models the quark-meson coupling
should also depend on the scalar soliton field. Similar conclusions
are reached by Krein, {\it et al.} \cite{Kre91}. However, we have
found that this causes unwanted behavior within the mean field
and Wigner-Seitz approximations used here (see Sec. 4), 
and so we report
calculations that use $\sigma$-independent quark-meson couplings.
As we shall see, this choice has the advantage of reproducing the
Quantum Hadrodynamics equation of state at low densities.

Thus we take the Lagrangian density to have the form
\begin{eqnarray} \label{Lagrange}
{\cal L} & = &\bar{\psi} \left[ i\gamma^{\mu} \partial_{\mu}-m_f
        -(g(\sigma) + g_s \phi - g_v \gamma^{\mu} V_{\mu})
          \right] \psi 
        + \frac{1}{2}\partial_{\mu}\sigma\partial^{\mu}\sigma - U(\sigma)
        \nonumber \\ & &	
\qquad        + \frac{1}{2}\partial_{\mu}\phi\partial^{\mu}\phi 
        - \frac{1}{2}m_{s}^2 \phi^2 
        - \frac{1}{4}F_{\mu\nu}F^{\mu\nu} 
        + \frac{1}{2}m_{v}^2 V_{\mu}V^{\mu}
\end{eqnarray}
where the $\sigma$ field self interaction is assumed to be
\begin{equation} \label{U(s)}
  U(\sigma) = \frac{a}{2!}\sigma^2
            + \frac{b}{3!}\sigma^3
            + \frac{c}{4!}\sigma^4 + B\; \; .
\end{equation}
The constants $a$, $b$ and $c$ are fixed so that $U(\sigma)$ has a
local minimum at $\sigma=0$ (inflection point if $a$=0)
and a global minimum at $\sigma=\sigma_v$,
the vacuum value. The mass of the glueball excitation
associated with the $\sigma$ field
is given by $m_{GB}=\sqrt{U''(\sigma_v)}$.

The quark-$\sigma$ coupling is taken to be 
\begin{equation} \label{g(s)} 
  g(\sigma) = 
  \left\{
  \begin{array} {cl}
  {\displaystyle{ g_{\sigma}\sigma }              }
   & \mbox{for the FL model } \\
  {\displaystyle{ g_{\sigma}\sigma_v 
  \left[\frac{1}{\kappa(\sigma)}-1\right]}} 
   & \mbox{for the $\chi$CDM } 
  \end{array}   \right. 
\end{equation}
where the chromodielectric function $\kappa(\sigma)$ has the form
\begin{equation} \label{kappa(s)}
 \kappa(\sigma)=1+\theta(x)x^{n_\kappa}[n_{\kappa}x-(n_{\kappa}+1)]
  \; \; ; \; x=\sigma/\sigma_v \; ,
\end{equation}
In the following we will take $n_{\kappa}=3$.
In perturbative calculations that include gluons
the dielectric function $\kappa(\sigma)$ is
regularized in order to handle infinities in the one gluon exchange
diagrams associated with a vanishing dielectric constant \cite{Fai88}.
Such a regulation is also useful for our numerical work.
Koepf et. al. \cite{Koe94} use the prescription
\begin{equation} \label{regkap}
  \kappa(\sigma) \longrightarrow \kappa(\sigma)(1-\kappa_v)+\kappa_v
\end{equation}
where $\kappa_v$ is a constant often simply fixed at $\kappa_v = 0.1$.
However, this regulation forces $g^\prime(\sigma_v)=0$, and we
prefer instead to regulate as follows:
\begin{equation} \label{kappa_R(s)}
 \kappa(\sigma)=1+\theta(x)x^{n_\kappa}[n_{\kappa}x-(n_{\kappa}+1-\kappa_v)].
\end{equation}
We find that properties of isolated solitons are independent of
$\kappa_v$ for values even as large as .1, but results become
sensitive to $\kappa_v$ at high densities.
Here, we shall treat $\kappa_v$ as an additional parameter. 

The Euler-Lagrange equations corresponding to (\ref{Lagrange})
are given by
\begin{equation} \label{EL1}
  \left[ \gamma^{\mu} \left(i\partial_{\mu}+ g_v V_{\mu} \right)
        - \left( m_f+ g(\sigma) - g_s \phi \right)
          \right] \psi = 0
\end{equation}
\begin{equation} \label{EL2}
    \partial_{\mu}\partial^{\mu}\sigma+U'(\sigma)
    + g'(\sigma)\bar{\psi} \psi = 0
\end{equation}
\begin{equation} \label{EL3}
     \partial_{\mu}\partial^{\mu}\phi + m_{s}^2 \phi 
     - g_s \bar{\psi} \psi = 0
\end{equation}
\begin{equation} \label{EL4}
  -\partial_{\mu} F^{\mu\nu} - m_{v}^2 V^{\nu} +
   g_v \bar{\psi} \gamma^{\nu}\psi = 0
\end{equation}
where $U'(\sigma) = \frac{d U(\sigma)}{d \sigma} $ and 
$g'(\sigma) = \frac{d g(\sigma)}{d \sigma}$ .
We solve Eqs. (\ref{EL1}-\ref{EL4}) in the mean field
approximation:  we replace the soliton field
$\sigma$ by a {\it c}-number
$ \sigma \longrightarrow \sigma(\vec{r})$
and the meson fields by 
their expectation values in the nuclear medium
 $ \phi \longrightarrow  < \phi > = \phi_0 $ and
 $ V_{\mu} \longrightarrow  < V_{\mu} > = \delta_{\mu 0} V_0 $,
with $\phi_0$ and $V_0$ constants.
The approximation that the scalar and vector mesons fields can be regarded
as constants while the soliton field is to depend on spatial coordinates
stems from the long range nature of the light mesons and the short range
nature of the soliton field due to the large glueball mass. In essence,
the mesons are fast degrees of freedom and the glueball is slow, and
we use a Born-Oppenheimer approximation.
The resulting equations for the quark and the scalar soliton field are,
\begin{equation} \label{MF1}
  \left[-i \vec{\alpha}\cdot \vec{\nabla} + g_v V_0 
        + \beta \left( m_f+ g(\sigma) - g_s \phi_0  \right)
          \right] \psi_k = \epsilon_k \psi_k
\end{equation}
\begin{equation} \label{MF2}
    -\nabla^2 \sigma + U'(\sigma)
    + g'(\sigma) \sum_{k(valance)}
       \bar{\psi_k} \psi_k = 0 \; .
\end{equation}
We consider only valence quarks in our calculations.

\section{The Wigner-Seitz approximation}
\label{sec:WS}

In order to use Eqs. (\ref{MF1}-\ref{MF2}) for the study of high 
density nuclear matter, it is sufficient to concentrate on a unit
cell and solve these equations with the appropriate boundary conditions.
In this paper we shall assume that each unit cell contains a single
nucleon, that is, our unit cell coincides with a Wigner-Seitz cell. 
The single nucleon energy $E_N$ is a sum of two terms, the energy
of the $n_q = 3$ quarks
and the energy carried by the scalar soliton field $\sigma$:
\begin{equation} \label{E_N}
  E_N = n_q \epsilon_q
      + \int_{\tiny WS \;\; cell} d \vec{r} 
        \left[ \frac{1}{2} (\vec{\nabla}\sigma)^2 + U(\sigma) \right] \; .
\end{equation}
The quark energy $\epsilon_q$ should be regarded as the eigenvalue of
Eq. (\ref{MF1}) for isolated bags, or as the average energy in the band
for dense nuclear matter. 
It should be noted that $E_N$ cannot be identified 
with the nucleon mass as it 
contains spurious center of mass motion \cite{Wil89}. In order to
correct the center of mass motion in the Wigner-Seitz cell the nucleon
mass at rest is taken to be
\begin{equation} \label{M_N}
  M_N = \sqrt{E_N^2-<P_{cm}^2>_{WS}} \; ,
\end{equation}
where $<P_{cm}^2>_{WS}=n_q <p_q^2>_{WS}+m_{GB}<(\vec{\nabla}\sigma)^2>_{WS}$.
The notation $<>_{\footnotesize WS}$ stands for an average over the 
Wigner-Seitz cell. Thus
$<p_q^2>_{WS}$ 
is the expectation value of the quark momentum squared
and $m_{GB}<(\vec{\nabla}\sigma)^2>_{WS}$ is the scalar soliton
momentum squared. The latter average is obtained using a
coherent state approximation \cite{Wil89}, which is essentially
a single mode correction to the classical soliton mass. We find this
latter correction to be small with respect to the total mass of
the soliton and ignore it henceforth (it is not clear how valid this
single-mode approximation is: at higher density we find that this correction
exceeds the energy of the $\sigma$ field, with both going to
zero).

At low density the band width vanishes and the quarks are
confined in separate bags. Then we expect the following
approximation to be most accurate: we assume the individual
nucleons move around as a gas of fermions with effective mass
$M_N$ given by Eq. (\ref{M_N}), and so
we get the following estimate for the total energy density at
nuclear density $\rho_B$:
\begin{equation} \label{Eeos}
  {\cal E} = \frac{\gamma}{2 \pi^2} \int_0^{k_F} 
      dk k^2 \sqrt{M_N^2+k^2} + \frac{1}{2} m_s^2 \phi_0^2
      - \frac{1}{2} m_v^2 V_0^2,
\end{equation}
where $\gamma=4$ is the spin-isospin degeneracy of the nucleons.
The Fermi momentum of the nucleons is related to the baryon density
through the relation
\begin{equation}\label{k_F}
	\rho_B=\frac{\gamma}{6 \pi^2} k_F^3 \; \; .
\end{equation}
The total energy per baryon is given by $E_B = {\cal E} / \rho_B $.
The constant scalar meson field $\phi_0$ is determined 
by the thermodynamic demand of minimizing ${\cal E}$, which
gives
\begin{equation} \label{minE}
  \phi_0 = -\frac{\gamma}{4\pi^2 m_s^2} 
 \int_0^{k_F} dk k^2 \frac{{d\,\over d\phi_0}(E_N^2-<P_{cm}^2>)}
{\sqrt{E_N^2-<P^2_{cm}>_{WS} + k^2}} .
\end{equation}
The vector meson field is determined by averaging the Euler-Lagrange
equation, Eq. (\ref{EL4}), on a Wigner-Seitz cell yielding
\begin{equation} \label{MF4}
     V_0    = \frac{g_v}{m_v^2} < \psi^{\dagger} \psi >
            = \frac{g_v}{m_v^2} \sum_{k(valance)}
              < \psi_k^{\dagger} \psi_k >_{\footnotesize WS}
              \rho_B 
            = \frac{n_q g_v}{m_v^2} \rho_B\; .
\end{equation}
These equations are similar to those of quantum hadrodynamics,
the difference here being that the nucleon now has structure and
thus the meson fields couple to the nucleon through its quarks.
At low density the nucleon mass approaches its free value, and
our mean field equations reduce to those of quantum hadrodynamics.


The approximation used here for the nuclear system has often
been adopted in modeling soliton matter, and is generally known as
the Wigner-Seitz approximation.
Each soliton is enclosed in a sphere
of radius $R$ such that 
$ \frac{4 \pi}{3}R^3=1/\rho_B $. 
On a periodic lattice the quark functions should satisfy Bloch's
theorem 
$\psi(\vec{r}+\vec{a})=e^{i \vec{k}\cdot \vec{a}}\psi(\vec{r})$ for the
lattice vectors $\vec{a}$. Concentrating on a single cell the Bloch theorem
gives boundary conditions 
for the quark spinors in that cell. Although one
can solve these boundary condition in a self consistent manner \cite{Web97},
we shall make the simplifying assumption of
identifying the bottom of the
lowest band by the demand that the derivative of the upper component of the
Dirac function disappears at $R$, and the top of that band by the demand 
that the value of the upper component is zero at $R$ \cite{Bir88}.
The quark spinor in the lowest
band is assumed to be an {\it s}-state 
\begin{equation} \label{q_spinor}
  \psi_k = \left( \begin{array} {c}
	          u_k(r) \\ i \sigma \cdot \hat{r} \; v_k(r)
                  \end{array} 
           \right) \chi,
\end{equation}
and the resulting Euler-Lagrange
equations for the spinor components are
\begin{equation} \label{u_r}
 \frac{d u_k}{d r} + \left[ m_f + g(\sigma) - g_s \phi_0 +
                    (\epsilon_k + g_v V_0) \right] v_k = 0
\end{equation}
\begin{equation} \label{v_r}
 \frac{d v_k}{d r} + \frac{2 v_k}{r}
                   + \left[ m_f + g(\sigma) - g_s \phi_0 - 
                    (\epsilon_k + g_v V_0) \right] u_k = 0 \; .
\end{equation}
The corresponding equation, (\ref{MF2}), for the soliton field assumes
the form
\begin{equation} \label{sigma_r}
    -\nabla^2 \sigma+U'(\sigma)
    + g'(\sigma) \rho_s(r)  = 0. 
\end{equation}
The quark density $\rho_q$ and the quark scalar density $\rho_s$
are given by
\begin{equation} \label{rho_q}
\rho_q(r) = \frac{n_q}{4 \pi \overline{k}^3 / 3 } \int_{0}^{\overline{k}} 
       d^3 k \; \left[ u_k^2(r) + v_k^2(r) \right] \; ,
\end{equation} 
\begin{equation} \label{rho_s}
\rho_s(r) = \frac{n_q}{4 \pi \overline{k}^3 / 3 } \int_{0}^{\overline{k}} 
       d^3 k \; \left[ u_k^2(r) - v_k^2(r) \right] \; ,
\end{equation} 
where the band is filled up to $\overline{k}$.
The quark functions are normalized so that there are
three quarks in the Wigner-Seitz cell.
The boundary conditions for the soliton field are 
$\sigma'(0)=\sigma'(R)=0$. The boundary conditions for the quark functions
at the origin are given by $u(0)=u_0$ and $v(0)=0$, where $u_0$ is determined
by the normalization condition
\begin{equation}\label{norm}
  \int_0^R 4 \pi r^2 d r (u(r)^2+v(r)^2) = 1 \; .
\end{equation} 
The boundary conditions at $r=R$ are given by
\begin{equation} \label{BCbot}
	u'_b(R)=0 \; \Rightarrow \; v_b(R)=0 
\end{equation}
for the bottom of the lowest band, and
\begin{equation} \label{BCtop}
	u_t(R)=0 
\end{equation}
for the top of the band.
Using these equations we can solve for the corresponding $\epsilon_b$
and $\epsilon_t$. We assume the tight-binding dispersion relation
\begin{equation} \label{Disp} 
\epsilon_k = \epsilon_b + (\epsilon_t - \epsilon_b)
\sin^2\left(\frac{\pi k}{2 k_t}\right), 
\end{equation}
and that the band is filled right 
to the top $k_t$. The assumption of such a dilute filling
has been made previously\cite{Wil89}, 
and we do not discuss it further. The quark
functions corresponding to the energy $\epsilon_k $ can be simply obtained
by integrating Eqs. (\ref{u_r}) and (\ref{v_r}) for each intermediate
value $\epsilon_k $.
Substituting the dispersion relation into Eq. (\ref{E_N}), the nucleon
energy is given by
\begin{equation} \label{E_Nr}
  E_N = \frac{3n_q}{k_t^3} \int_0^{k_t} dk k^2 
        \epsilon_k
      + \int_{0}^{R} 4 \pi r^2 d r 
        \left[ \frac{1}{2} \sigma'(r)^2 + U(\sigma) \right] \; ,
\end{equation}
which is then used in (\ref{M_N}) to determine the equation of
state (\ref{Eeos}).

At higher density, where the bags begin to
overlap, there is no reason to assume each quark
is tightly bound to a single bag nor to impose that 3$q$ groups move
collectively with a well-defined momentum. Our equation of
state cannot then be considered a good approximation at
higher densities, and we must then find a
different approximation for handling the kinetic energy.
This is the subject of the next paper in this series. For
now we are content with studying low density behavior, with
a special interest in whether our approximations can still
produce saturation at the expected density. In fact, as one approaches
the empirical nuclear saturation density, the different models
begin to distinguish themselves. This is discussed in detail in the
next section.

\section{The Trivial solution }
\label{sec:trivial}
	
The NTS models we are considering have a uniform
plasma phase that is preferred at high densities.  This
corresponds to the solution $\sigma=0$, so that the soliton
bags ``dissolve'' and the quarks are free. For the original FL model,
this solution is favored at unreasonably low densities. 
Moreover, in the WS calculations \cite{Wil89}, 
for cell radii
below $\approx$0.8-0.9 fm only a trivial constant solution can be
found. For the more 
sophisticated analysis reported in \cite{Web97}, which includes
higher partial waves in the quark wave functions and direct
computation of the bands, the calculation
breaks down already at a cell radius $\approx$1.4 fm. 
We view this as a fault of the model and not of the WS
approximation, for regardless of whether our restricting the soliton
to a single spherical cell represents a dense system well or not,
our model of the nucleon should nevertheless allow the soliton to be 
squeezed to volumes well below nuclear density before it
breaks apart. (The saturation density of nuclear matter is
$\rho_0$=0.17fm$^{-3}$, which corresponds to a WS cell
radius $R_0$=1.12fm.) 
This leads us to look for improvements upon
the original FL model, and the two particular extensions studied
in this paper are generalizations of the quark-glueball coupling
$g(\sigma)$ and the addition of explicit meson degrees of
freedom.
Before proceeding to the numerical results, let us look
at how these extensions affect the trivial solution. 

The trivial solution to the WS equations (\ref{u_r}-\ref{sigma_r})
is $u_k(r)=u_0$, $v_k(r)=0$ and $\sigma(r)=\sigma_0$, where the
constants $u_0$ and $\sigma_0$ are independent of both $r$ and
$k$. Strictly, for $k>0$ this solution does not satisfy our
boundary conditions; however, we find that below
a given cell volume the numerical solution develops a singularity
at the boundary in order to reproduce this preferred solution.
Moreover, the more accurate self-consistent calculations of
\cite{Web97} for the FL model (which do not assume just $s$-wave
states in the quark wave function) show that the width of the
band also narrows sharply at the onset of the trivial solution,
indicating that the $k$-dependence is not important. Thus in
order to find the trivial solution that
characterizes the breaking down of the WS calculation,
we may assume that all the quarks are at the bottom of the band.
This corresponds to squeezing an isolated soliton, and we
insist that our model favors a nontrivial solution until the
cell radius gets well below that corresponding to nuclear
density.

In the present section we shall consider altering the quark-meson
couplings in our models as suggested in Lagrangian \cite{Nie82,Ban97},
namely:
\begin{equation}
g_s \to g_s g(\sigma), \qquad g_v \to g_v g(\sigma).
\end{equation}
This will allow us to investigate whether the coupling to meson
fields can help cure the breaking down of the FL model at too
low density.

For the trivial solution, then, 
the normalization condition (\ref{norm}) gives
\begin{equation}
u_0 = \left({3\over 4\pi R^3}\right)^{1/2} = \rho_B^{1/2}.
\end{equation}
From Eq. (\ref{v_r}) we find the quark eigenenergy is
\begin{equation} \label{eps_tr}
\epsilon = m_f + g(\sigma_0)[1 - g_s \phi_0
+ g_vV_0],
\end{equation}
where $\sigma_0$ obeys Eq. (\ref{sigma_r}), modified to
account for the change in $q$-$\sigma$ coupling, which leads to
\begin{equation} \label{sig_tr}
U'(\sigma_0) = -3\rho_B g'(\sigma_0)[1-g_s\phi_0+g_vV_0].
\end{equation}
Solving for the mean field values of
$\phi_0$ and $V_0$, we have
\begin{equation} \label{phi_tr}
\phi_0 =  {3g_s\over m_s^2}\rho_B g(\sigma_0)
\qquad {\rm and} \qquad
V_0 = {3g_v\over m_v^2} \rho_B g(\sigma_0),
\end{equation}
yielding finally (for $m_f$=0)
\begin{equation} \label{eps2_tr}
\epsilon = g(\sigma_0)\left\{1 - 3\rho_B g(\sigma_0)
\left[{g_s^2\over m_s^2}-{g_v^2\over m_v^2}
\right]\right\}
\end{equation}
and the total energy density
\begin{equation} \label{E_tr}
{\cal E} = 3 \rho_B 
g(\sigma_0) \left\{ 1 - {3\over 2}\rho_B
g(\sigma_0)\left[{g_s^2\over m_s^2}-{g_v^2\over m_v^2}
\right]\right\} + U(\sigma_0).
\end{equation}
This is clearly a spurious solution corresponding to putting
all the quarks in the lowest level. One needs to add the
kinetic energy correctly to get the true energy of the
quark plasma.

We would like to select a model for which the trivial solution
is found only at densities much higher than nuclear density, and
we would like the quark mass $\epsilon$ to be zero in the
preferred phase. This latter condition implies $\sigma_0$=0.
However, for the FL model $g(\sigma)=g_\sigma \sigma$,
and therefore $\sigma_0$=0 is not a solution
of Eq. (\ref{sig_tr}). Using Eqs. (\ref{phi_tr}) and
substituting for $U', g'$ and $g$, for the
FL model Eq. (\ref{sig_tr}) becomes
\begin{equation} \label{sig2_tr}
a\sigma_0 + {b\over 2}\sigma_0^2 + {c\over 6}\sigma_0^3
 = -3\rho_B g_\sigma \left[1
-3 \rho_B g_\sigma
\left({g_s^2\over m_s^2}-{g_v^2\over m_v^2}\right) \sigma_0 \right].
\end{equation}
If we turn off the meson fields $g_s,g_v\to 0$, we see that
the solution to (\ref{sig2_tr}) goes as $-\rho_B^{1/3}$
at large densities. Not 
only does this blow up as $\rho_B\to\infty$, but it gives the
quarks an unphysical negative mass 
$\epsilon \sim -g_\sigma \rho_B^{1/3}$. 
On the other hand, if we turn on the meson fields and make the usual
choice of parameters so that $g_s^2/m_s^2>g_v^2/m_v^2$ (which
is necessary for saturation), there is now a positive
high density solution that goes as $\rho^{-1}_B$. However,
there is still an unphysical negative solution, now
diverging as $\rho_B$. Thus we do not expect the inclusion of
the meson fields to cure the problem of the FL model.

Now let us consider a modified FL model in which we
take the quark-glueball coupling to be $g(\sigma)=g_\sigma \sigma^2$.
We shall call this the FL$^2$ model. Then instead of
Eq. (\ref{sig2_tr}) we have
\begin{equation} \label{FL2}
a\sigma_0 + {b\over 2}\sigma_0^2 + {c\over 6}\sigma_0^3
 = -6\rho_B g_\sigma \sigma_0 \left[1
-3 \rho_B g_\sigma
\left({g_s^2\over m_s^2}-{g_v^2\over m_v^2}\right) \sigma_0^2 \right].
\end{equation}
For this model the trivial solution $\sigma_0=0$ exists in the
Wigner-Seitz approximation. There are also nonzero solutions,
however. With the meson fields switched off there are two
nontrivial solutions
\begin{equation}
\sigma_0 = -{3b\over 2c}\left\{1 \pm \sqrt{1-{8c\over 3b^2}
(a + 6\rho_B g_\sigma)}\right\},
\end{equation}
which exist only at densities $\rho_B < a(3b^2/8ac - 1)/6g_\sigma$.
For the parameter choice used here, this corresponds to cell
radii $R > 1.67$ fm. These solutions give the quarks a positive
mass and do not necessarily signal any problems with the model.
For nonzero $q$-$\sigma$ coupling, the solutions exist
also at high density, behaving as $\sigma_0 \propto \pm\rho_B^{-1/2}$.
The quark effective mass (\ref{eps2_tr}) vanishes
in this limit. Thus the FL$^2$ does not seem to have the problems
of the FL model when using the WS approximation.

Finally, let us consider the $\chi$CD models. From
Eqs. (\ref{g(s)}) and  (\ref{kappa(s)}) we see that for
small $\sigma$ the coupling $g(\sigma)$ is of leading
order $n_\kappa$ in $\sigma$. This means that, regardless
of whether the mesons are present, for the $\chi$CD model
with $n_\kappa\ge 2$
there will be a solution to Eq. (\ref{sig_tr}) such that
$\sigma_0=0$. This corresponds to the desired free massless quark
plasma phase, with vanishing meson field averages, since
then $g(0)=g'(0)=0$. Furthermore, one can show that as
$\rho_B\to\infty$ the only other solution has $\sigma_0\sim
\rho_B^{-1}$
and gives the quarks a positive mass that also vanishes
in the high density limit. Thus we expect the $\chi$CD
models to provide a more reasonable description of dense
nuclear matter than does the FL model, for there is no
unphysical trivial solution to the WS equations that
will cause problems.

Indeed, the type of behavior predicted by our analysis of
the trivial solution is reflected in our Wigner-Seitz
calculations. 
As we show in Fig. 1, the $\chi$CD model exhibits
Wigner-Seitz solutions down to very small cell radii (we never
reached a breakdown point, taking $R$ as low as
$0.05$fm), whereas the FL model breaks
down at the expected point $R\approx 0.8$fm.  
The FL$^2$ model can be taken
down to lower cell radii than the FL, breaking down
at $R\approx.25$fm, well above the transition to the
uniform plasma phase. However, the behavior of $\epsilon_b$ below
$R=1$fm for the FL$^2$ model does not seem desirable. This could
be cured by allowing the quark-meson coupling to depend on
$\sigma$, but we find then non-smooth transitions to new non-trivial
solutions in all the NTS models discussed here. 
We find more reasonable results if we keep the quark-meson coupling
independent of $\sigma$, and thus
the $\chi$CD seems the preferable model. We shall present results only
for this latter model in the next section.

\section{Results }
\label{sec:results}

We have performed the calculations using a straightforward
numerical integration of the equations of motion for the
$q$ and $\sigma$ fields. The mean field $\phi_0$ is found
by directly minimizing the free energy (\ref{Eeos}). 
As a check,
we have also performed some calculations with a relaxation
routine, but this is far more time consuming.
The numerical calculations with the $\chi$CD model
were carried out using
the following set of parameters 
\begin{equation}
 a=50 \, \mbox{fm$^{-2}$ }, \; b=-1300 \, \mbox{fm$^{-1}$ }, \; 
 c=10^4, \;  g_{\sigma}=2, \; \kappa_v = 0.1 \; .
\end{equation}
The parameters of the potential $U(\sigma)$ are chosen to
give a reasonable bag constant $B = 46.6$MeV/fm$^3$ and
glueball mass $m_{GB} = 1.82$GeV.
The value of $\sigma_v$ is calculated to be $.285$fm$^{-1}$, the 
single soliton energy is $1391$ MeV and the nucleon mass is $1176$ MeV.
The nucleon rms radius is $0.876$ fm. The parameters were chosen
to fit the rms radius and nuclear matter properties, resulting in
a somewhat high value of the nucleon mass. We did not spend much
effort in fine-tuning the parameters, however.
As the short range repulsion provided by the vector field 
in QHD is provided by the Wigner-Seitz boundary conditions in 
the liquid soliton model, in what follows we shall take the coupling
constant between the vector meson and the quarks to be zero, $g_v=0$.
On the other hand we shall treat $g_s$ as a free parameter in 
order to study its effect on the soliton matter.

For completeness, we also list the parameter sets used for the other
models studied in the previous section. For the FL model, we take
the parameters as in Birse et. al. \cite{Bir88}. (This set was
also used in Ref. \cite{Web97}.)
The full set of parameters for the soliton model is,
\begin{equation}
 a=0.0 \; , b=-700.43 \, \mbox{fm$^{-1}$ } 
       \; , c=10^4 \; , g_{\sigma}=10.98 \; .
\end{equation}
These parameters correspond to a single soliton energy $1260$ MeV,
and nucleon mass $902$ MeV after removing the center of mass motion. 
For the FL2 model we use the same potential parameters and
a coupling $g_\sigma = 60$fm$^{-1}$. 

We proceed to our presentation of the results for the
$\chi$CD model. To our knowledge, this is the first application
of this particular model to the study of dense matter. Previous
work has dealt with few nucleon systems \cite{Wil89}.

In Figs. 2 and 3 we show the fields $u(r)$ and $\sigma(r)$ for
several values of the cell radius $R$. As can be seen from its
behavior at the cell boundary, the quark field $u_k(r)$
displayed in Fig. 2 is that corresponding to the bottom of
the band $k=0$. Fig. 3 shows that the depth of the bag decreases
only slighty as the cell radius is lowered, indicating that the
quarks remain tightly bound. (Note that for the FL model, however,
the bag depth actually increases with smaller $R$ \cite{Bir88,Wil89}.)
That the quarks are tightly bound in the bag can be seen clearly
in Fig. 4, where we display the quark density $\rho_q(r)$ given
by Eq. (\ref{rho_q}). There is only a small overlap with quarks
from neighboring cells even at radii below 1 fm. In Figs. 1-4,
the quark-meson coupling has been set to zero.

In Fig. 5 the top and bottom of the lowest energy band is shown
as a function of $R$ for several values of the quark-meson
coupling $g_s$. The point at which the band begins to form,
$R\approx 1.6$fm, is not very sensitive to the value of $g_s$.
On the other hand, the structure of the band depends strongly on
the coupling to the scalar meson, becoming wider as
$g_s$ is increased. In Fig. 6 we present
the energy of the soliton $E_N$ given by Eq. (\ref{E_Nr}), and
in Fig. 7 we show the total energy per baryon $E_B$ of the
system, derived from Eq. (\ref{Eeos}).
As expected, the introduction of the scalar meson to the
the soliton matter results in attraction and saturation. 
The empirical nuclear
saturation density corresponds to $R_0=1.12$ fm, whereas from
Fig. 7 we see, for example, that our NTS nuclear matter 
energy per baryon has a minimum at $R\approx 1.35$fm
for $g_s=1$. The saturation energy is $E_B - M_N^{(as)} \sim
20$ MeV, as compared to the empirical value $E_{sat}=16$ MeV.
The compression modulus of nuclear matter is
$K=R^2 d^2E_B/dR^2$ at the equilibrium point: for the
present model we find $K\approx 1170$MeV. This is to be
compared to empirical estimates that usually lie in
the range $100$-$500$MeV, with $K=200$MeV the generally
accepted value\cite{M92}.
These results should be regarded as only order of magnitude,
as at higher density we are surely underestimating  the
kinetic energy of the system. We shall develop a more reasonable
model of the liquid state, which leads to a more quantitatively accurate
EOS for solitonic nuclear matter, in the next paper of this
series.

The solution for the $\chi$CD model is stable down to
very low values of $R$ (at least as low as $0.05$ fm), and it is apparent that
with proper calibration this model can be used as an excellent starting
point for the study of nuclear matter.
Another important feature of this model is the increase of the nucleon
rms with density, Fig. 8. This unexpected outcome of the model is
in accord with the EMC effect. Checking
the sensitivity of this effect to model parameters, we found that 
the increase in the rms is insensitive to recoil corrections, but
might disappear with different choices of the soliton parameters.

\section{Discussion}
\label{sec:con}

In this paper we have investigated a class of non-topological soliton
models that generalize the Friedberg-Lee model. We have studied
nuclear matter in the Wigner-Seitz approximation, where neighboring
bags begin to overlap at higher density, as a means of distinguishing
between the various models. The models differ in the precise form
of the coupling between the quarks and the scalar gluon field, and
in the presence and form of couplings to explicit meson fields.
Of the models studied, we have found the chiral chromodielectric 
model to exhibit behavior best in line with phenomenological
expectations. For this model, there is no breakdown in the Wigner-Seitz
calculation as the cell radius is decreased, thus overcoming  a serious
shortcoming of the original FL model.
We consider it a computational necessity to consider only constituent
quarks and ignore colored gluons and sea quarks, and thus
these restrictions along with the addition of explicit meson fields may 
be seen as an approximation to the original theory.
The original $\chi$CD model is closely based upon QCD, exhibiting
absolute confinement, and it is therefore satisfying to see that
this model is selected by our studies of nuclear matter.

We include mesons along the lines of quantum hadrodynamics, and we
use the mean field approximation.
In our final calculations we have considered only the scalar meson,
but it is straightforward to include also the vector meson, which
should be done for fine-tuning the parameters. 
We note that we found it best to use a quark-meson coupling that
is not modulated by the quark-glueball coupling, in contrast to
that argued for in Refs. \cite{Kre91,Ban97}. Indeed, when using a $q-\phi$
coupling proportional to $g(\sigma)$, we find jumps to qualitatively
different solutions as density is increased. The resulting curve of
energy as a function of cell radius has several bumps, which is
clearly an undesirable feature. However, this should not necessarily
be taken as an argument against models that employ quark-meson
couplings that depend on $\sigma$, but rather merely a statement
that when using the mean field approximation it is more consistent
to use a $\sigma$-independent $q$-$\phi$ coupling $g_s$. Indeed, one
can view this coupling as a ``mean field approximation'' to a
more fundamental coupling --- namely,
 $g_s \sim \langle \tilde{g}_s g(\sigma)\rangle$.

In the end, then, we have used a $\chi$CD model with $\sigma$-independent
coupling between quarks and the scalar meson. We find that the
inclusion of the scalar meson provides a clear saturation point
for nuclear matter and that a rough fit to both empirical nuclear
matter and single nucleon properties can be obtained. One of the
most interesting features found is an increase in the nucleon rms
radius at intermediate densities, in line with the EMC effect.
This is dependent on the presence of the scalar meson. There are
clearly several refinements to our model and our calculations that
must be considered before fine-tuning the parameters. Among these
are the inclusion of other mesons in the model, the addition
of perturbative gluonic effects, and an improved
calculation of the quark wave function along the lines of Ref. \cite{Web97}.
One can also improve our handling of the spurious CM motion.
Foremost, however, we need to improve our modeling of the
liquid state.

In this paper we have used a low-density approximation in modeling
nuclear matter with solitons. This consists in using a Wigner-Seitz
approximation to calculate an effective nucleon mass. Then the kinetic
energy is added to the system by taking the motion of the nucleons to
be that of a Fermi gas. The assumptions of this picture include: 
the nuclear medium restricts any given nuclear
bag to a spherical cell; 
any given nucleon moves slowly, so that it can 
be constructed at rest and then boosted; 
the quarks remain tightly bound inside the bag; 
the nucleons move independently within the medium.
Only the first of these assumptions can remain valid at
high densities, where the nuclear medium will form a liquid
in which each nucleon's motion is localized on short time
scales and quarks need not remain tightly bound inside
the bag. Since we are ultimately interested in studying
the transition to a uniform quark plasma within our
non-topological soliton model, we must do a better
job of modeling the liquid state at high densities.
This is the subject of our next paper.

\section*{Acknowledgments} 
 
We thank J.A. Tjon for helpful discussions.

\pagebreak

\begin{figure}[htb]
\caption{\protect\label{fig1}
Energy of the bottom level of the band as a function of
cell radius for the Friedberg-Lee (FL), modified
Friedberg-Lee (FL$^2$) and Chiral Chromodielectric ($\chi$CDM)
 models}
\hskip-2.3cm
\includegraphics{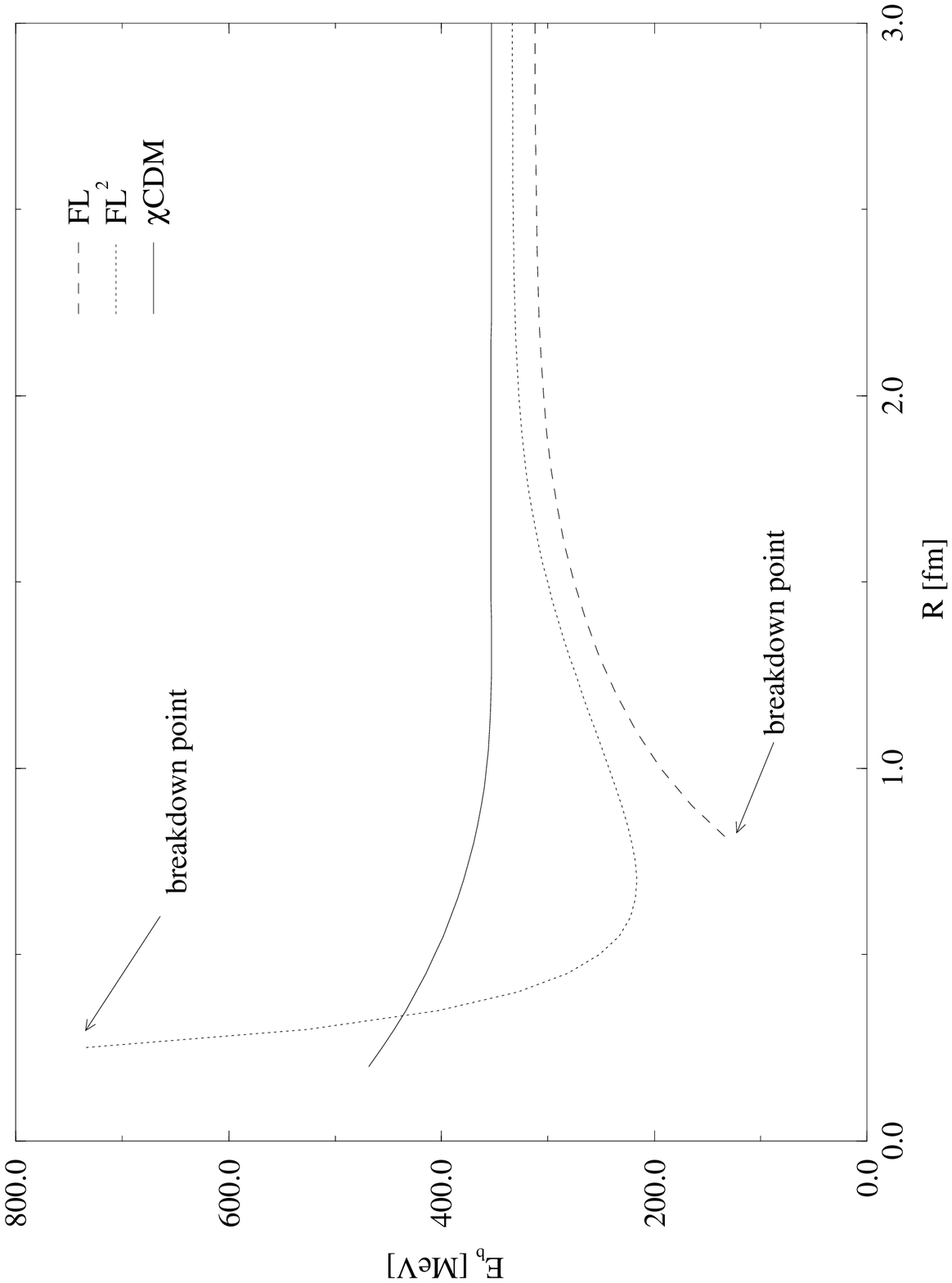}
\vskip8.0cm
\end{figure}

\,
\newpage

\begin{figure}[htp]
\caption{\protect\label{fig2}
Upper component $u(r)$ of quark wave function at the bottom of
the band for several cell radii $R$ in $\chi$CD model}
\hskip-1.4cm 
\includegraphics{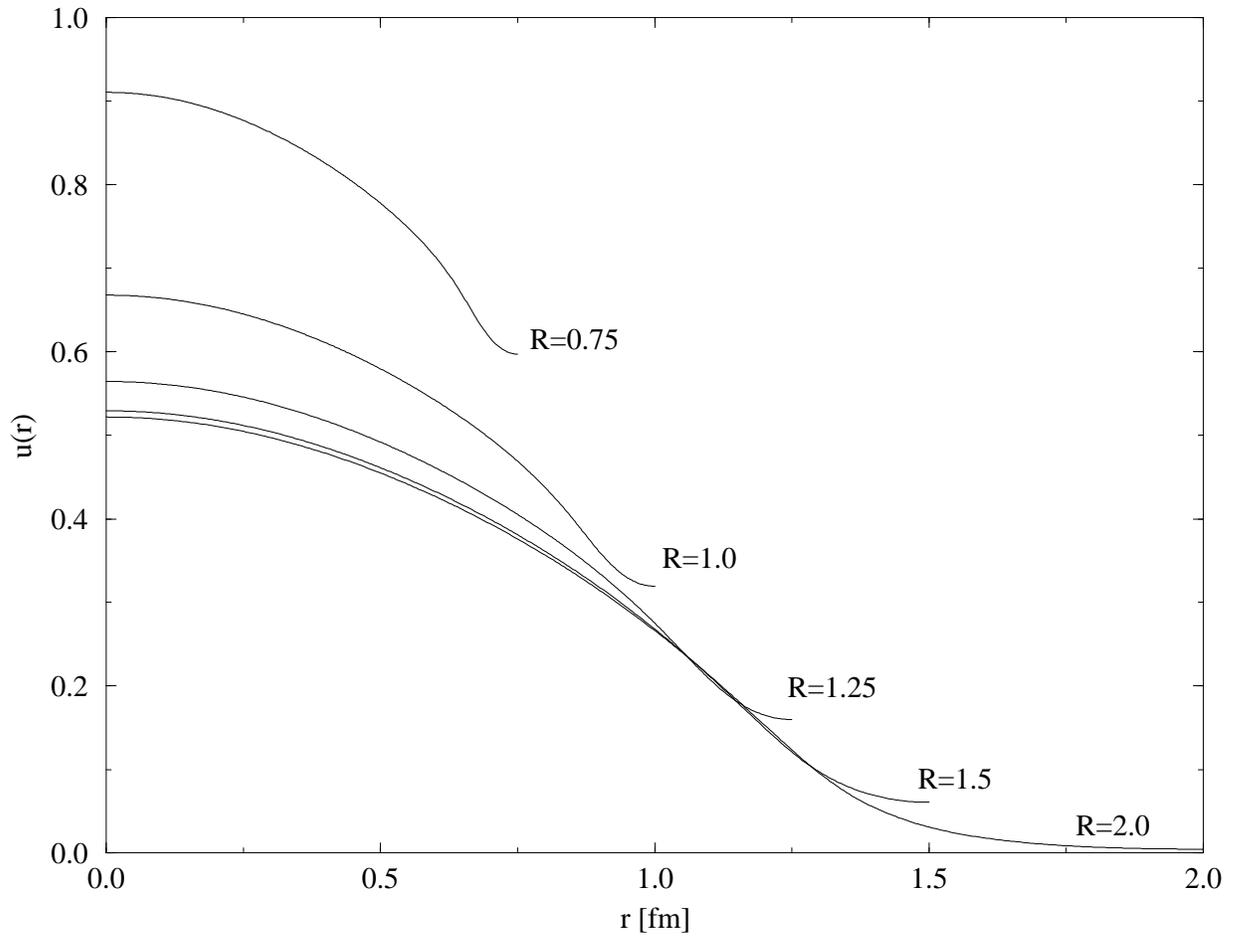} 
\vskip-3.0cm
\end{figure}

\newpage

\begin{figure}[htp]
\caption{\protect\label{fig3}
Soliton bag $\sigma(r)$ for several cell radii $R$ in
$\chi$CD model}
\hskip-2.3cm
\includegraphics{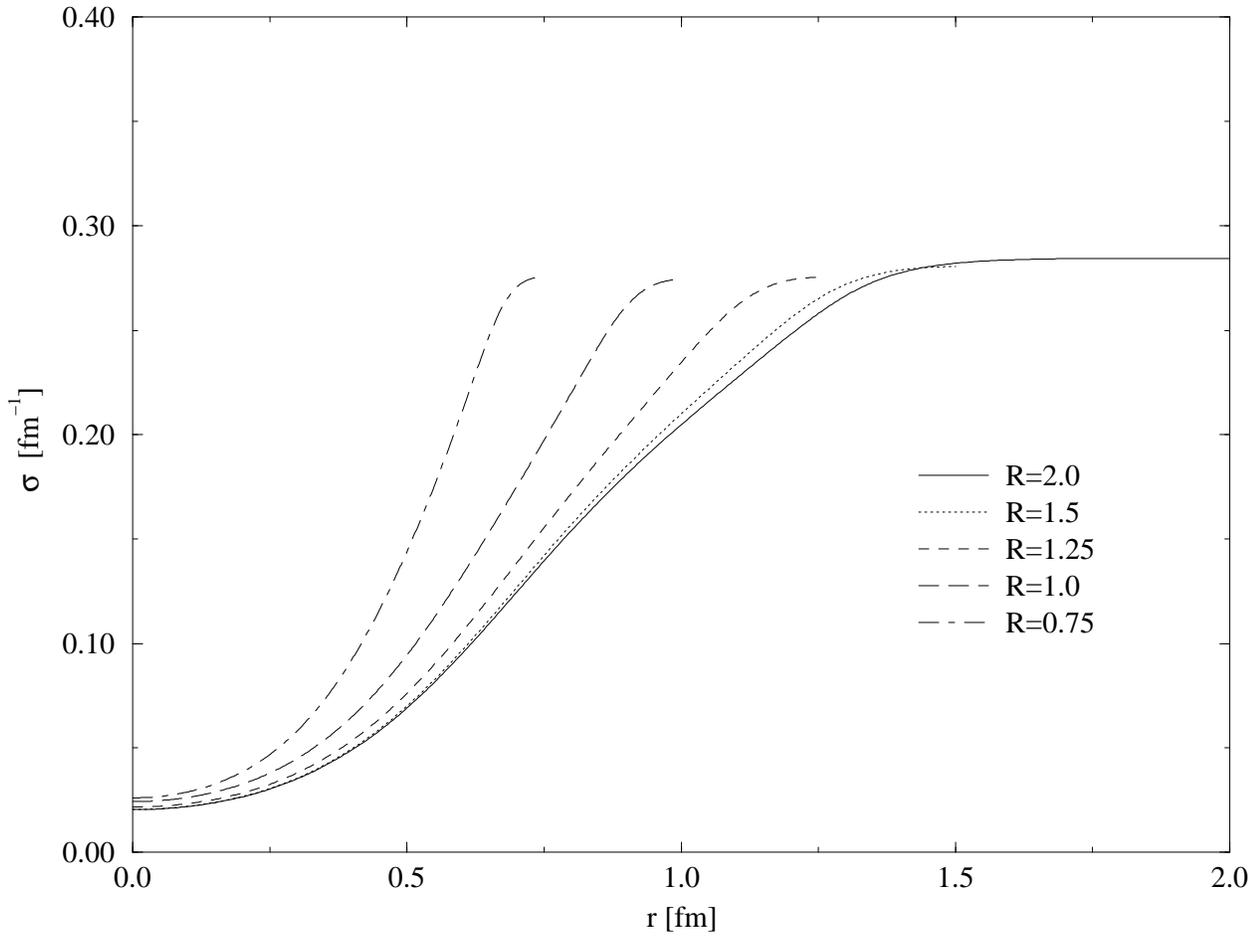}
\end{figure}

\newpage

\begin{figure}[htp]
\caption{\protect\label{fig4}
Quark density $\rho_q(r)$ for several values
of $R$ in $\chi$CD model}
\hskip-2.2cm 
\includegraphics{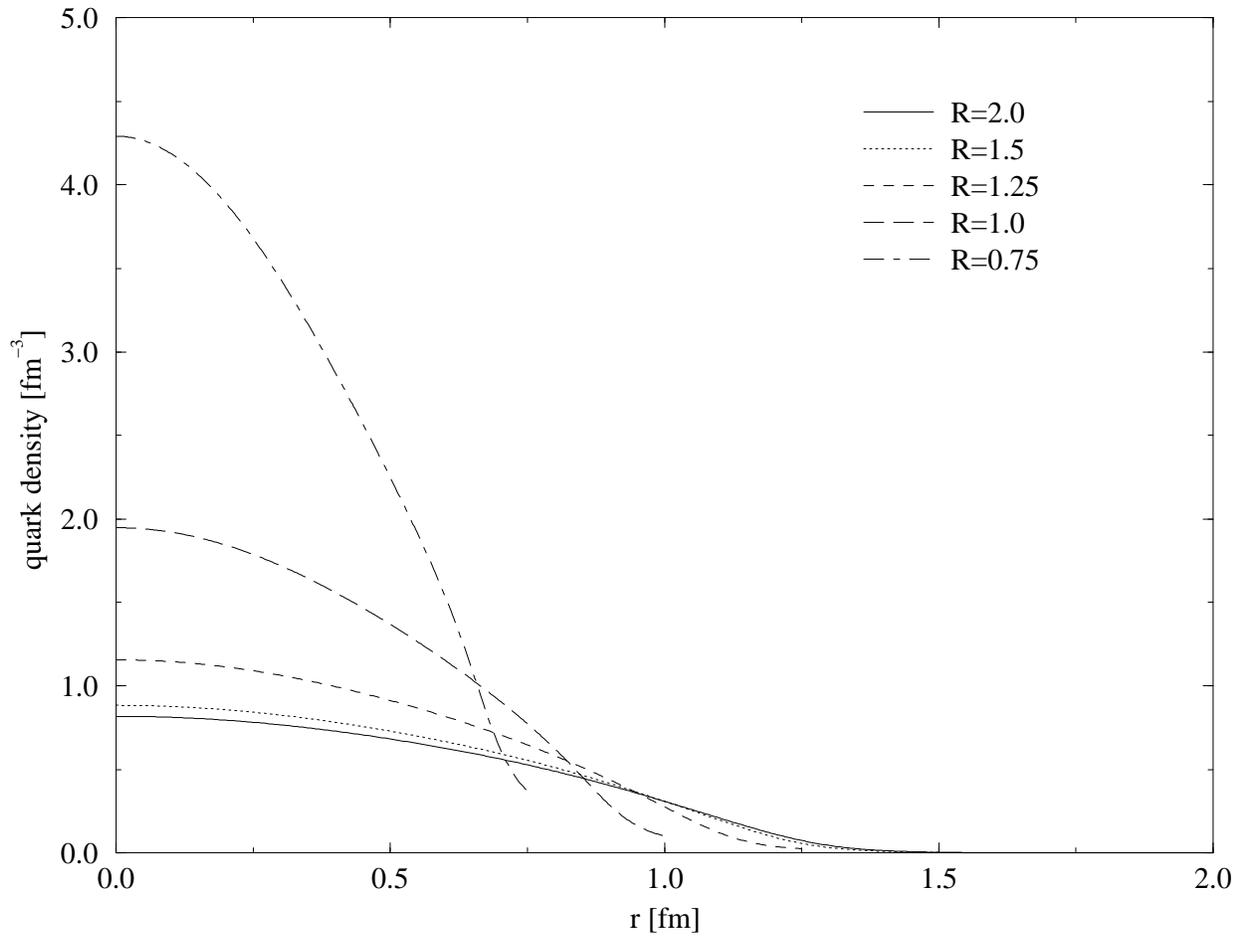} 
\vskip-3.0cm
\end{figure}

\newpage

\begin{figure}[htp]
\caption{\protect\label{fig5}
Energy of the bottom and top levels of the band
versus $R$ for several values of the quark-meson
coupling $g_s$}
\hskip-1.6cm 
\includegraphics{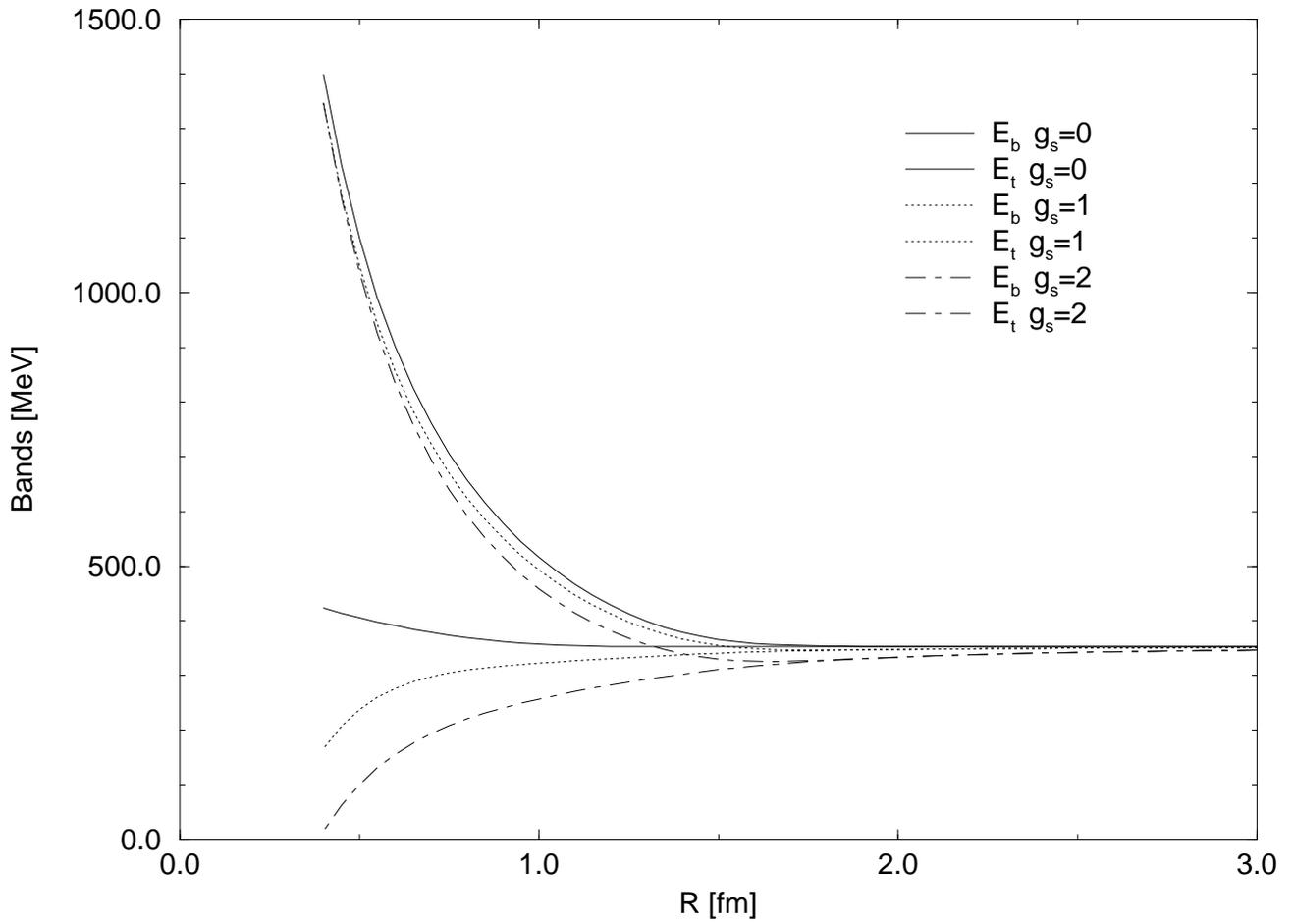} 
\vskip-3.0cm
\end{figure}

\newpage

\begin{figure}[htp]
\caption{\protect\label{fig6}
Soliton energy $E_N$ (before subtracting spurious CM motion)
for several values of $g_s$}
\hskip-1.4cm 
\includegraphics{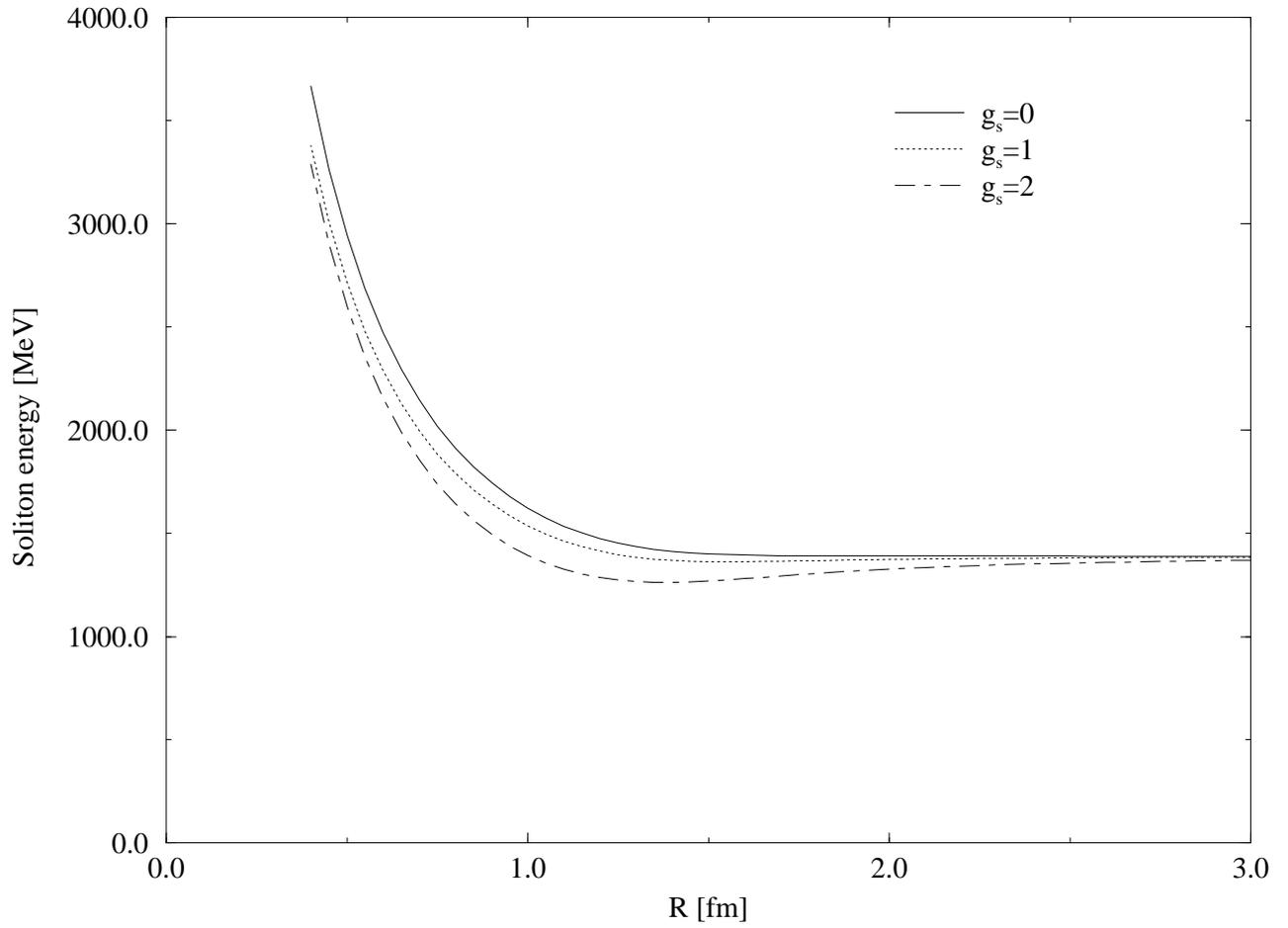} 
\vskip-3.0cm
\end{figure}

\newpage

\begin{figure}[htp]
\caption{\protect\label{fig7}
Energy per baryon $E_B$ of nuclear matter as function
of $R$ and $g_s$ in the $\chi$CD model}
\hskip-1.8cm 
\includegraphics{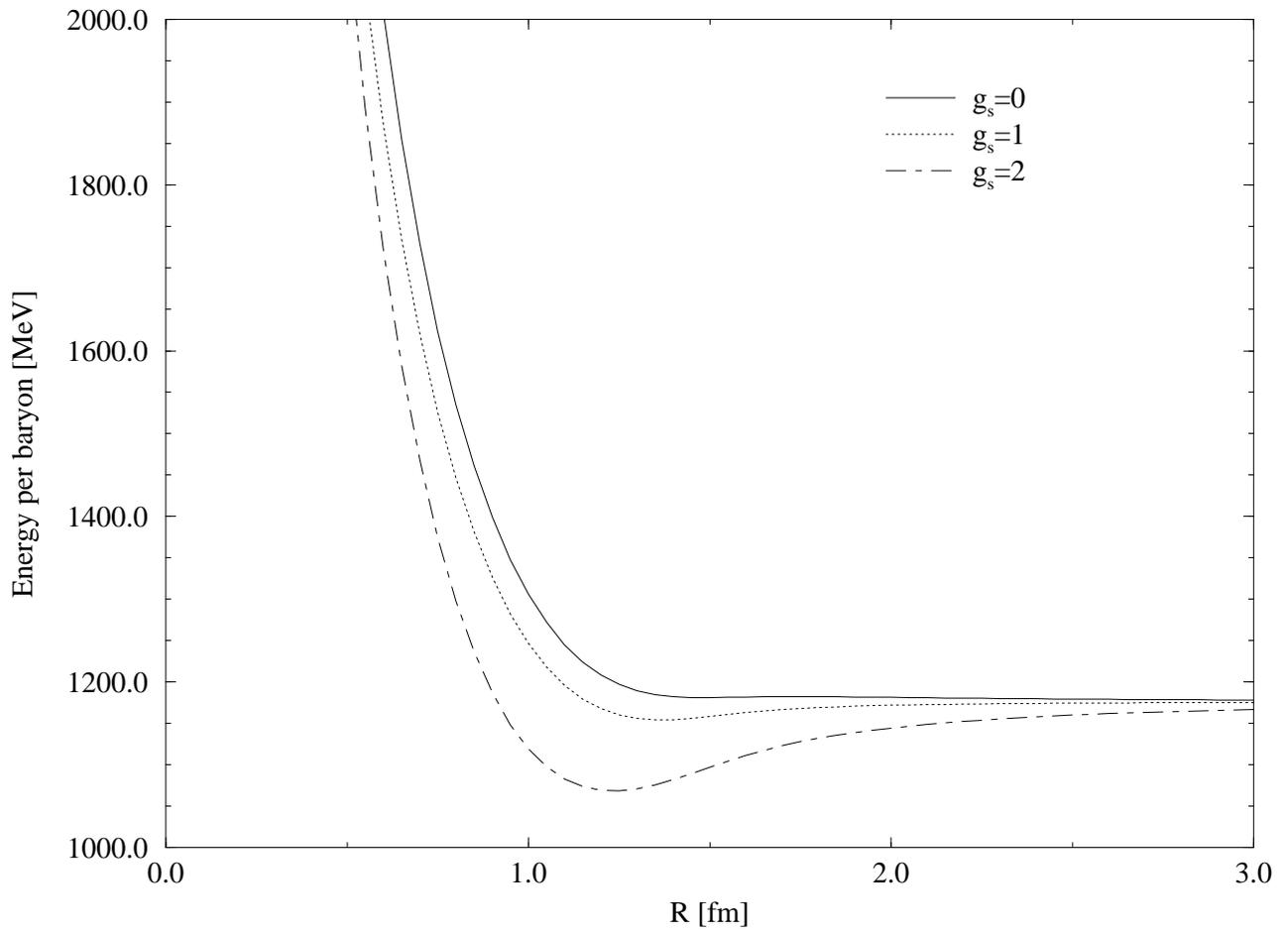} 
\vskip-3.0cm
\end{figure}

\newpage

\begin{figure}[htp]
\caption{\protect\label{fig8}
Nucleon rms radius as function of $R$ and $g_s$
in the $\chi$CD model}
\hskip-1.8cm 
\includegraphics{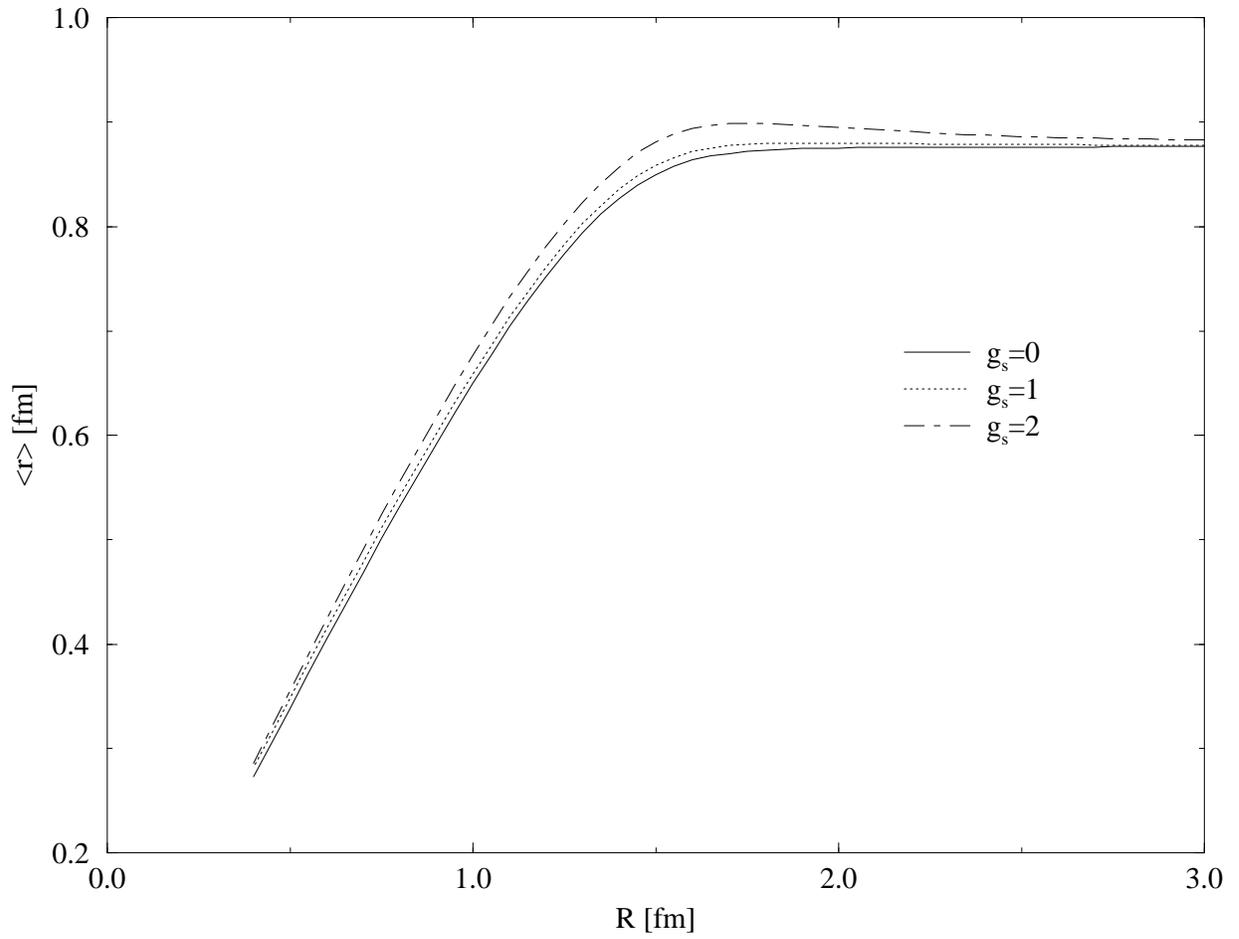} 
\vskip-3.0cm
\end{figure}

\newpage

\end{document}